\documentclass[10pt,letterpaper,twocolumn]{article}

\usepackage{ol2}
\usepackage[draft]{hyperref}
\usepackage{amsmath}
\usepackage{graphicx}
\usepackage{amssymb}

\begin{document}

\twocolumn[

\title{Robust laser frequency stabilization by serrodyne modulation}

\author{Ralf Kohlhaas,$^{1,*}$ Thomas Vanderbruggen,$^1$ Simon Bernon,$^{1,\dagger}$ Andrea Bertoldi,$^1$ Arnaud Landragin,$^2$ and Philippe Bouyer$^{1,3}$}
\address{$^1$ Laboratoire Charles Fabry, UMR 8501, Institut d'Optique, CNRS, Univ. Paris Sud 11, Campus Polytechnique, \\
2 Avenue Augustin Fresnel, F-91127 Palaiseau cedex, France\\
$^2$ LNE-SYRTE, Observatoire de Paris, CNRS and UPMC, 61 avenue de l'Observatoire, F-75014 Paris, France\\
$^3$ Laboratoire Photonique, Numerique et Nanosciences - LP2N Universit\'e Bordeaux - IOGS - CNRS : UMR 5298 \\
Bat A30, 351 cours de la liberation, Talence, France\\
$^*$Corresponding author: Ralf.Kohlhaas@institutoptique.fr\\
$^\dagger$Now at Universit\"at T\"ubingen, D-72076 T\"ubingen, Germany}

\begin{abstract}

We report the relative frequency stabilization of a distributed feedback erbium-doped fiber laser on an
optical cavity by serrodyne frequency shifting. A correction bandwidth of 2.3 MHz and a dynamic range of 220
MHz are achieved, which leads to a strong robustness against large disturbances up to high frequencies. We
demonstrate that serrodyne frequency shifting reaches a higher correction bandwidth and lower relative
frequency noise level compared to a standard acousto-optical modulator based scheme. Our results allow to
consider promising applications in the absolute frequency stabilization of lasers on optical cavities.

\end{abstract}


]

\noindent The frequency stabilization of lasers is required in a wide range of applications such as optical
atomic clocks \cite{Jiang2011}, fiber sensors \cite{Chow2005}, gravitational wave detectors
\cite{Zawischa2002}, and quantum optomechanical setups \cite{Verlot2010}. In these systems, the correction
bandwidth and correction range are two key parameters to reach a low noise stabilization as well as to ensure
the robustness against perturbations from the environment. Nevertheless, many commonly used lasers, such as
fiber, dye or diode-pumped solid state lasers, have only piezo-electric transducers (PZT) as a means for
frequency correction. This limits the correction bandwidth to typically a few kHz. As a consequence, for
higher frequencies, an external actuator is usually needed to extend the correction range. An acousto-optical
modulator (AOM) can only reach a few hundred kHz bandwidth and a dynamic range of up to a few tens of MHz
\cite{Bertoldi2010}. Instead, an electro-optical modulator (EOM) allows a higher correction bandwidth of
several MHz but frequency shifts cannot be sustained and it operates on a small correction range
\cite{Hall1984,Ohmae2010}. An optimal frequency actuator should pair the high bandwidth of an EOM with a large
dynamic range and have the ability to hold the correction signal.

In this letter we present the implementation of optical serrodyne frequency shifting \cite{Wong1982} in a stabilization
scheme. This opens the way to use an EOM for a large correction bandwidth together with a broad correction range. It is
thus very robust since it can compensate for large and fast disturbances. Moreover, the system offers very low optical
losses thanks to the high shifting efficiency, and the method is suitable for all-fibered or integrated optics setups.

\begin{figure}[b]
\centerline{\includegraphics[width=7.5cm]{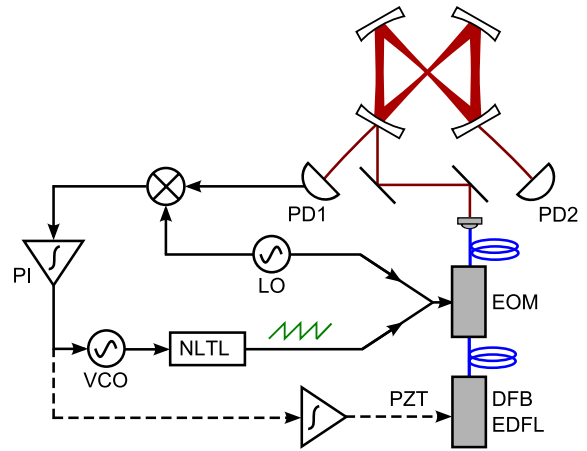}}
\caption{(Color online) Setup of the laser stabilization on an optical cavity based on serrodyne frequency shifting (see text for notations). An optional feedback path (dashed lines) is added on the piezo transducer.}
\label{fig:1}
\end{figure}

Serrodyne frequency shifting consists in the phase modulation of an optical wave with a saw-tooth signal of frequency
$f_{\rm saw}$ and phase amplitude $2\pi m$ $(m \in \mathbb{N})$, leading to a frequency displacement of $m f_{\rm saw}$.
The required saw-tooth waveform can be generated by a nonlinear transmission line (NLTL), a passive component which
transforms a sinusoidal waveform in a high fidelity saw-tooth signal with the same fundamental frequency. Recently,
serrodyne modulation was applied by feeding the output of a NLTL into an EOM. This lead to serrodyne frequency shifts from
200 MHz to 1.6 GHz and efficiencies as high as 80$\%$ \cite{Houtz2009,Johnson2010}. By changing the frequency of the
sinusoidal wave at the input of the NLTL, the frequency of the light at the output of the EOM is tuned. This can be used as
an actuator in laser frequency stabilization.

\begin{figure}[t]
\centerline{\includegraphics[width=8.1cm]{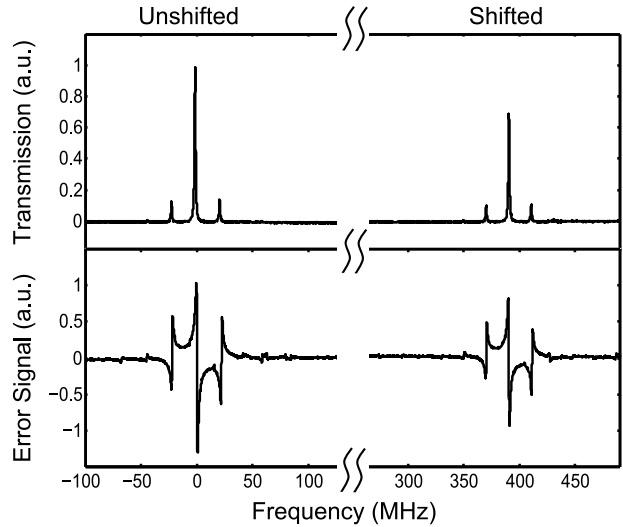}}
\caption{(Top) The transmission signal obtained on PD2 when scanning the length of the optical cavity without (left) and with (right) serrodyne frequency shifting. (Bottom) The error signal obtained by the Pound-Drever-Hall technique is also shifted by the serrodyne modulation.}
\label{fig:2}
\end{figure}

The method is demonstrated by the stabilization of a laser on an optical cavity, presented in Fig. \ref{fig:1}. The
employed cavity has a butterfly configuration \cite{Bernon2011}. It has a finesse of 1788 at a wavelength of 1560 nm and a
free spectral range of 976 MHz. A distributed feedback erbium-doped fiber laser (DFB EDFL, Koheras, NKT Photonics) at 1560
nm with a typical linewidth of a few kilohertz is injected into the cavity. The correction signal is obtained with the
Pound-Drever-Hall (PDH) technique \cite{Drever1983}. The optical beam is phase modulated at the local oscillator (LO)
frequency of 20 MHz, and is detected with an InGaAs photodiode (PD1) after reflection on the injection mirror of the
cavity. The demodulation with the LO provides an error signal (Fig.~\ref{fig:2}) which is sent to a proportional-integrator
(PI) controller. The correction signal at the output of the controller is summed to a voltage offset chosen to set the open
loop operating frequency of the voltage controlled oscillator (VCO, ZX95-625-S+, Minicircuits) to 390 MHz. The output of
the VCO is amplified to a power of 27 dBm by a rf amplifier (ZHL-1-2W-S-09-SMA, Minicircuits) to reach the optimal
serrodyne shifting efficiency. The signal feeds a NLTL (7112-110, Picosecond Pulse Labs, 300-700 MHz nominal input range)
which generates harmonics of the fundamental frequency up to 20 GHz. The resulting saw-tooth waveform is then combined to
the LO signal using a broadband power combiner (ZX10R-14-S+, Minicircuits) and sent to the fibered EOM (PM-0K5-00-PFA-PFA,
Eospace). The amplitude of the saw-tooth was chosen to match a phase shift of 2$\pi$ on the phase modulator, thus leading
to a frequency shift of the light equal to the frequency of the VCO. In a thermally noncontrolled environment, large long
term drifts may overcome the correction range of the serrodyne scheme. In such a case, a low frequency correction can be
added on the laser piezo transducer (PZT) by an additional integration of the output of the PI (see Fig. \ref{fig:1}). This
additional correction also keeps constant the mean frequency of the serrodyne modulation and therefore the serrodyne
shifting efficiency.

In Fig.~\ref{fig:2}, the transmission of the optical cavity and the error signal are shown when the length of the resonator
is scanned. It demonstrates the feasability to produce the frequency shifting and the modulation sidebands on a single EOM.
The incident light and the error signal are shifted by 390 MHz. A shifting efficiency of 69~$\%$ is reached on the
transmission, whereas the shifting efficiency of the PDH signal is 76~$\%$. Since the EOM is already included in the
optical system for the error signal generation, an additional optical loss of only 1.6~dB is introduced by the serrodyne
modulation. Indeed, light which is not frequency shifted remains either at the initial frequency (about 3~$\%$ of the total
power) or is transferred to higher harmonics of the modulation frequency (28~$\%$). These spurious frequency components are
intrinsically filtered by the optical cavity. The frequency shifting range goes from 280 MHz (limited by the VCO) to 500
MHz (limited by the rf amplifier before the NLTL), corresponding to a span of 220 MHz. It could be potentially extended to
more than 1 GHz using another NLTL (e.g. NLTL 7113-110, Picosecond pulse labs), associated with a suitable controllable
oscillator and rf amplifier.

\begin{figure}[b]
\centerline{\includegraphics[width=8.1cm]{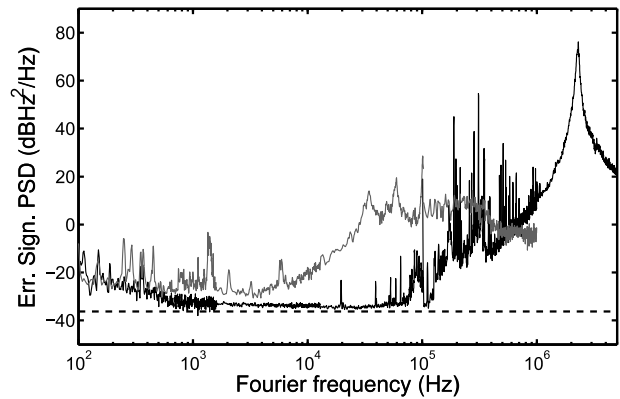}}
\caption{Noise PSD of the error signal with the serrodyne technique (black) and the stabilization with an AOM in double pass (grey). The dashed line represents the light shot noise limit on PD1.}
\label{fig:3}
\end{figure}

With the serrodyne shifting technique the stabilization of the laser to the optical cavity with a bandwidth of  2.3 MHz is
obtained. This bandwidth is limited by phase shifts occuring in the servo electronics. The noise power spectral density
(PSD) of the error signal with the laser locked to the cavity is shown in Fig.~\ref{fig:3}, where it is compared with that
obtained using a double pass AOM actuator on the same laser and cavity system \cite{Bertoldi2010}. It was measured with a
FFT spectrum analyzer, then converted to relative frequency noise using the slope of the PDH signal. In addition, it was
corrected for the transfer function of the cavity of $(1+(2f/f_{0})^2)^{-1}$ where $f_{0}$~=~546~kHz is the cavity
linewidth (FWHM).  The servo bump at 2.3 MHz indicates the bandwidth of the lock. The noise spectrum is limited from 2 kHz
to 50 kHz by the light shot noise (within 3 dB) and below 2 kHz by the voltage noise of the input operational amplifier of
the PI (THS4601, Texas Instruments).  When compared with the noise PSD previously obtained with the double pass AOM system
\cite{Bertoldi2010}, the lower noise level from 1 kHz to 500 kHz indicates the advantage of the higher correction bandwidth
provided by serrodyne frequency shifting.

\begin{figure}[b]
\centerline{\includegraphics[width=7.9cm]{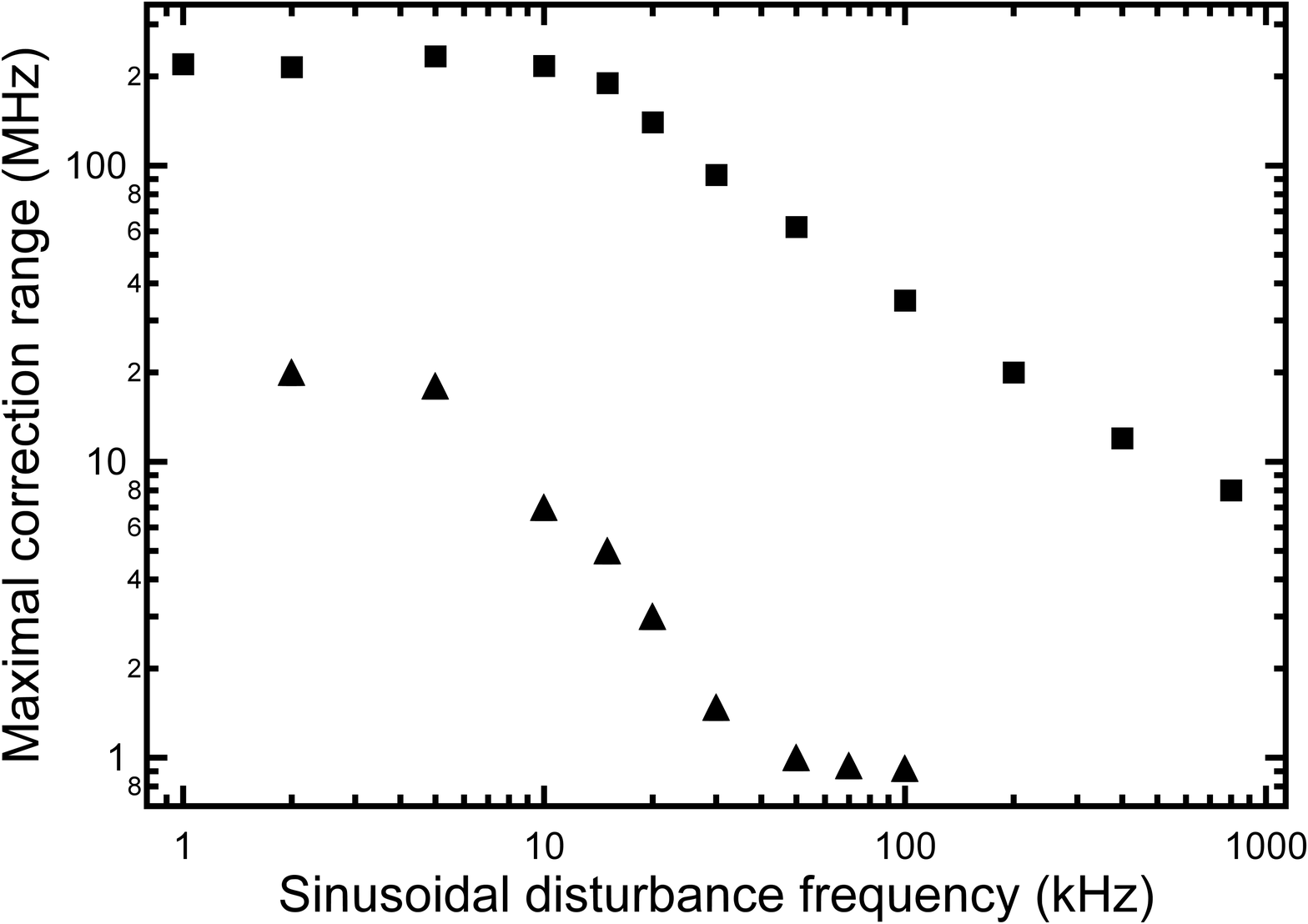}}
\caption{Comparison of the lock robustness obtained with the serrodyne shifting technique (squares) and an AOM system (triangles).}
\label{fig:4}
\end{figure}

The high loop bandwidth of 2.3 MHz and the broad dynamic range of 220 MHz lead to a very robust stabilization
against large disturbances. To simulate perturbations, we add a sinusoidal waveform to the correction signal
that drives the VCO. In addition, we define the maximal correction range as the point beyond which the
external perturbation leads to frequency fluctuations between the laser and the cavity equal to the cavity
linewidth. The value of this limit versus the perturbation frequency is shown in Fig.~\ref{fig:4}. From 1 kHz
to 10 kHz, the upper limit is given by the full dynamic range. Above 10 kHz, the correction range is limited
by the gain of the PI controller. For perturbations at 800 kHz the system can still correct for frequency
shifts of 8 MHz. The robustness with serrodyne modulation is compared with the one of the double pass AOM
system [6]. Here, a dynamical range of 20 MHz and a correction bandwidth of 250 kHz are found, as typical for
AOM based systems. The serrodyne stabilization loop thus allows to correct for frequency disturbances which
are at least one order of magnitude larger than for an AOM and thanks to its higher bandwidth it can act where
an AOM does not operate. In both systems the PZT transducer will increase the maximal correction range for
frequencies below a few hundred Hz.

We demonstrated the relative frequency stabilization of a laser on an optical cavity by serrodyne frequency shifting. The
technique benefits from an EOM for frequency correction which leads to a high correction bandwidth (2.3 MHz here). At the
same time, a large correction range can be achieved (220 MHz here). Since the EOM is already included for the error signal
generation, only electronical components are added to implement the feedback loop. This leads to a simple and efficient
frequency stabilization scheme, which could be further improved by electro-optical integration techniques
\cite{Johnson2011}.  We show by comparison with an AOM based system that the higher bandwidth achieved with the serrodyne
modulation leads to a lower relative frequency noise level. This can be exploited in the absolute frequency stabilization
of lasers on ultra stable cavities. In addition, serrodyne frequency shifting provides very robust locks. It opens the way
to use optical resonators in transportable systems which operate in harsh environments \cite{Geiger2011}, e.g.
for optical frequency generation \cite{Mimoun2008} or applications in atomic physics \cite{Bernon2011}. 

We thank F. Moron and A. Villing for technical assistance. We acknowledge funding support from DGA, IFRAF and EU (project
iSENSE). The work of A. B. was supported by the IEF Grant PIEF-GA-2009-235375.

\end{document}